\documentclass[secnumarabic, graphics,floatfix, nofootinbib,tightenlines,nobibnotes, aps, prl, 12pt]{revtex4-1}
\usepackage{graphicx}
\usepackage[english]{babel}
\usepackage[utf8]{inputenc}
\usepackage{amsmath,amssymb}
\usepackage{amsfonts}
\usepackage{multirow}

\begin{document}

\title{Hydrodynamic approach to the centrality dependence of di-hadron correlations}

\author{Wagner M. Castilho$^{1}$}
\author{Wei-Liang Qian$^{2,1}$}
\author{Fernando G. Gardim$^{3}$}
\author{Yogiro Hama$^{2}$}
\author{Takeshi Kodama$^{4}$}

\affiliation{$^1$Universidade Estadual Paulista J\'ulio de Mesquita Filho, SP, Brazil}
\affiliation{$^2$Universidade de S\~ao Paulo, SP, Brazil}
\affiliation{$^3$Universidade Federal de Alfenas, MG, Brazil}
\affiliation{$^4$Universidade Federal do Rio de Janeiro, RJ, Brazil}

\date{April 13, 2017}

\begin{abstract}
Measurements of di-hadron azimuthal correlations at different centralities for Au+Au collisions at 200 AGeV were reported by the PHENIX Collaboration. 
The data were presented for different ranges of transverse momentum. 
In particular, it was observed that the away-side correlation evolves from double- to a single-peak structure when the centrality decreases. 
In this work, we show that these features naturally appear as due to an interplay between the centrality-dependent smooth background elliptic flow and the one produced by event-by-event fluctuating peripheral tubes. 
To compare with the PHENIX data, we also carry out numerical simulations by using a hydrodynamical code NeXSPheRIO, and calculate the correlations by both cumulant and the ZYAM method employed by PHENIX Collaboration.
It is shown that our results are in reasonable agreement with the data.
A brief discussion on the physical content of the present model and its difference from other viewpoint is also presented.

\end{abstract}

\pacs{PACS numbers: 25.75.Ld}

\maketitle
\newpage

\section{I. Introduction}

The di-hadron correlations in ultra-relativistic heavy-ion collisions are fundamental observables reflecting the properties of the underlying particle production mechanisms.
The observations of enhancement in correlations at intermediate and low $p_T$ \cite{star-ridge3,phenix-ridge5,phobos-ridge7,alice-vn4,cms-ridge7,atlas-vn4}, in comparison with those at high $p_T$ \cite{star-jet-1,star-jet-2}, strongly indicate the hydrodynamics/transport nature of such phenomena \cite{sph-corr-1,hydro-v3-1,ph-vn-3,ph-corr-1,hydro-corr-1,ampt-4,ampt-5} (in the former case).
The structure on the near side of the trigger particle is referred to as ``ridge''.
It features a narrow peak in $\Delta\phi$, located around zero, and a long extension in $\Delta\eta$, and therefore is related to a long-range correlation in pseudo-rapidity.
The away-side correlation broadens, in $\Delta\phi$, from peripheral to central collisions, and it exhibits double peak for certain centralities and particle-$p_T$ ranges.
The latter is usually called ``shoulders''.

On the theoretical side, the current understanding from the extensive studies of event-by-event basis (EbE) hydrodynamic analysis is that particle correlation for the transverse momentum less than those of jet domain can be mostly interpreted by collective flow, or to be specific, by the flow dynamics observed in terms of anisotropic parameters, $v_{n}$, through the hydrodynamic evolution.
It has been shown that these parameters are closely related to the corresponding $\epsilon_n$, the energy density anisotropy parameters of the initial conditions (IC) \cite{hydro-v3-1,hydro-v3-2}.
As a matter of fact, the observed behavior of anisotropic parameters as functions of centrality and transverse momentum can well be studied at a very high quantitative level \cite{hydro-v2-heinz-2,hydro-v2-voloshin-1} and the correlation between $v_{n}$ and $\epsilon_n$ has also been established (for more references see the recent reviews \cite{hydro-review-8,hydro-review-4,hydro-review-9,hydro-review-6}).
In this sense, the di-hadron correlations observed in PHENIX \cite{phenix-ridge5} are generally attributed to the triangular flow, created by the fluctuations in the IC.
However, in spite of its success of the EbE hydrodynamic simulations, the physical picture of the correlations among $v_n$ and $\epsilon_n$ in the hydrodynamic evolutional scenario become less clear for harmonics greater than $n=2$. 
In fact, it is pointed out that the correlation between $v_n$ and $\epsilon_n$ as functions of centrality and transverse momentum becomes weaker for larger harmonics, although their event average values are almost linearly correlated \cite{hydro-vn-3}. 
Accordingly, the values of  $v_3$ and $\epsilon_3$, for example, do not necessarily have one to one correspondence \cite{sph-vn-4}.
On the other hand, a real hydrodynamic event is a deterministic process, so that to understand the mechanism of how these EbE fluctuations are created, it is essential to clarify the initial and subsequent dynamics of relativistic heavy ion collisions on the real event by event basis.  
In this context, the peripheral tube model \cite{sph-corr-2,sph-corr-3,sph-corr-4,sph-vn-4} provides a very straightforward and reasonable picture for the generation of the triangular flow and consequently the di-hadron correlations within the usual EbyE hydrodynamic approach. 
There, the fluctuations in initial energy density distributions are separated into two parts; one the smooth elliptic distribution which is determined by the collision geometry, and the other, spatially localized sharp high-density spots (hot spots).  
High-energy tubes do also appear in the middle of hot matter in more realistic EbyE IC, but these are quickly absorbed by the surrounding matter (see Fig.5 of \cite{sph-corr-4}), so only peripheral tubes are relevant in our discussion.

In previous works \cite{sph-corr-ev-2,sph-corr-ev-4}, the peripheral tube model was employed to discuss the trigger-angle dependence of the di-hadron correlation in mid-central collisions (20\%-60\% centrality).
This approach seeks to provide an intuitive interpretation for the observed characteristics of the two particle correlations, rather than to replace the realistic event by event hydrodynamical calculations.
A key component of the model is that the background collective flow is deflected by a peripheral tube (or ``hot spot") emerging from the fluctuating initial conditions, and subsequently contributes to the resulting two-particle correlation.
In this context, the ``deflected flow" from the tube is not aligned with the event plane but associated with the azimuthal angle of random initial fluctuations.
In fact, this distinct feature offers a more straightforward description of the observed trigger-angle dependence of the di-hadron correlation \cite{sph-corr-ev-4}.
By a simplified analytical approach, the resultant correlations were understood as due to the interplay between two components, namely, the elliptic flow caused by the initial almond-shaped deformation of the whole system and the flow produced by tubelike fluctuations.
Specifically, the latter is already present in the central collisions, with the characteristic 3-peak shape of correlation in $\Delta\phi$ \cite{sph-corr-2,sph-corr-3,sph-corr-4}.
The contribution of the former is trigger-angle dependent: it is back-to-back (peaks at $\Delta\phi= 0 , \pi$) in the case of in-plane triggers ($\phi_s=0$) and it is shifted by $\pi/2$ (peaks at $\Delta\phi=-\pi/2 , \pi/2$) in the case of out-of-plane triggers ($\phi_s = \pi/2$).
Therefore, in the out-of-plane direction, the ``valley" at $\Delta\phi = \pi$ helps to form the observed double-peak structure; while in the in-plane direction, the peak at $\Delta\phi = \pi$ strengthens the correlation on away side, resulting in a single peak.
The present work follows this line of thought, we extend the above model to study the centrality dependence of the di-hadron correlation.
In order to discuss the average correlation at different centrality windows, we integrate the results obtained in \cite{sph-corr-ev-4} over the azimuthal angle of the trigger particle.
The centrality dependence comes out naturally from that of the background elliptic flow, whose magnitude increases from central to peripheral collisions.
As a result, when one goes from peripheral to central collisions, the away-side correlation is expected to evolve from single- to double-peak structure and meanwhile, the magnitude of the correlation increases.
This was exactly observed in the measurements carried out by PHENIX Collaboration \cite{phenix-ridge5}.

The present work is organized as follows. 
In section II, we show that the main feature of the two particle correlations can be qualitatively reproduced by using the peripheral tube model.
The numerical simulations are carried out in section III by employing the hydrodynamical code NeXSPheRIO, and the correlations are evaluated by both cumulant and the ZYAM method.
Conclusion remarks are given in the last section.

\section{II. The peripheral tube model}

The purpose of the peripheral-tube model is to show in a clear-cut way how several characteristics of the so-called ridge phenomena are produced \cite{sph-corr-ev-2,sph-corr-ev-4}. 
Being so, although extracted from the more realistic studies, for the sake of clarity, only essential ingredients are retained in the model. Namely,
\begin{itemize}
\item The collective flow consists of contributions from the background and those induced by randomly distributed peripheral tubes. 
For simplicity and clearness, here we consider just one such tube. 
The main difference, if one has more than one peripheral tube, is that in the latter case (even in the central collisions), the flow parameters ($v_n$ and $\Psi_n$) are largely spread event-by-event, 
although the final two-particle correlation is almost independent of the number of tubes (See Ref.\cite{sph-corr-ev-5,sph-corr-7}).
\item The background elliptic-flow coefficient increases from central to peripheral collisions, while the multiplicity decreases.
\item Event-by-event fluctuation is reflected in the model in two aspects: first, the azimuthal location of the tube is randomized from event to event and, second, the background multiplicity fluctuates from event to event.
\end{itemize}
Recall that experimentally the background multiplicity always fluctuates and this is important to be considered in the correlation calculation, as will become clear later.
We write down the one-particle distribution as a sum of two contributions: the distribution of the background and that of the tube.
\begin{eqnarray}
  \frac{dN}{d\phi}(\phi,\phi_t) =\frac{dN_{bgd}}{d\phi}(\phi) +\frac{dN_{tube}}{d\phi}(\phi,\phi_t),
 \label{eq1}
\end{eqnarray}
where
\begin{eqnarray}
 \frac{dN_{bgd}}{d\phi}(\phi)&=&\frac{N_b}{2\pi}(1+2v_2^b\cos(2\phi)),    \label{eq2}\\
 \frac{dN_{tube}}{d\phi}(\phi,\phi_t)&=&\frac{N_t}{2\pi}\sum_{n=2,3}2v_n^t\cos(n[\phi-\phi_t]).   \label{eq3}
\end{eqnarray}
For simplicity, we assume that, besides the radial flow, the background distribution in Eq.(2) is dominated by the elliptic flow, which is observed experimentally, especially for non-central collisions. In Eq.(\ref{eq2}), the flow is parametrized in terms of the elliptic flow parameter $v_2^b$ and the overall multiplicity, denoted by $N_b$. As for the contribution from the tube, we take into account the smallest possible number of parameters to reproduce the shape of the two-particle correlation due to a peripheral tube in an isotropic background energy distribution \cite{sph-corr-ev-2,sph-corr-ev-4}. Therefore, only two components $v_2^t$ and $v_3^t$ are retained in Eq.(\ref{eq3}). 
Owing to non-linear nature of hydrodynamics, the approximation in Eq.(1) is more reliable when the fluctuations are small, 
though we believe that the results drawn from the model remain qualitatively valid for more realistic cases.
We also note here that the overall triangular flow in our approach is generated only by the tube and so its symmetry axis is correlated to the tube location $\phi_t$.
The azimuthal angle $\phi$ of the emitted hadron and the position of the tube $\phi_t$ are measured with respect to the event plane $\Psi_2$ of the system. Since the flow components from the background are much bigger than those generated by the tube, as discussed below, $\Psi_2$ is essentially determined by the elliptic flow of the background $v_2^b$.
For the same reason, we prefer, in this analysis, not to include the radial-flow component in the tube contributions, so $N_b$ in Eq.(2) may be literally interpreted as the overall multiplicity.

Following the methods used by PHENIX Collaboration \cite{phenix-ridge5}, the subtracted di-hadron correlation is given by
\begin{eqnarray}
 \left<\frac{dN_{\text{pair}}}{d\Delta\phi}\right>   =\left<\frac{dN_{\text{pair}}}{d\Delta\phi}\right>^{\text{proper}} -\left<\frac{dN_{\text{pair}}}{d\Delta\phi}\right>^{\text{mixed}}.
 \label{eq4}
\end{eqnarray}
In peripheral tube model,
\begin{eqnarray}
 \left<\frac{dN_{\text{pair}}}{d\Delta\phi}\right>^{\text{proper}}
 =\int\frac{d\phi_s}{2\pi}\frac{d\phi_t}{2\pi}f(\phi_t)
 \frac{dN}{d\phi}(\phi_s,\phi_t)\frac{dN}{d\phi}
 (\phi_s+\Delta\phi,\phi_t),
\end{eqnarray}
where $f(\phi_t)$ is the distribution function of the tube and $\phi_s$ is the azimuthal angle of the trigger particle.
We will take $f(\phi_t)=1$, for simplicity.

The combinatorial background
$\left<{dN_{\text{pair}}}/{d\Delta\phi}\right>^{\text{mixed}}$
can be calculated by using either cumulant or the ZYAM method \cite{zyam-1,zyam-2}. As shown below, both methods lend very similar conclusions in our model.
Here, we first carry out the calculation using cumulant, which gives
\begin{eqnarray}
\left<\frac{dN_{\text{pair}}}{d\Delta\phi}\right>^{\text{mixed}\text{(cmlt)}}
 =\int\frac{d\phi_s}{2\pi}\frac{d\phi_t}{2\pi}f(\phi_t)
 \int\frac{d\phi_t'}{2\pi}f(\phi_t')\frac{dN}{d\phi}
 (\phi_s,\phi_t)\frac{dN}{d\phi}(\phi_s+\Delta\phi,\phi_t').
\end{eqnarray}
Notice that, in the averaging procedure above, integrations both over $\phi_t$ and $\phi_t'$ are required in the mixed events, whereas only one integration over $\phi_t$ is enough for proper events. This will make an important difference between two terms in the subtraction of Eq.(\ref{eq4}).

Using our simplified parametrization, Eqs.(\ref{eq1}-\ref{eq3}) and, by averaging over events, one obtains
\begin{eqnarray}
\left<\frac{dN_{\text{pair}}}{d\Delta\phi}\right>^{\text{(cmlt)}}
 &=&\frac{\langle N_b^2\rangle -\langle N_b\rangle ^2}{(2\pi)^2}\left(1+2(v_2^b)^2\cos(2\Delta\phi)\right) \nonumber\\
 &+&(\frac{N_t}{2\pi})^2\sum_{n=2,3}2({v_n^t})^2
 \cos(n\Delta\phi).
 \label{ccumulant}
\end{eqnarray}
Observe that the multiplicity fluctuation gives rise to a difference between the factors multiplying the background terms of the proper- and mixed-event correlations. 
Therefore, the background elliptic flow is {\it not} canceled out but does contribute to the correlation.
From the r.h.s. of Eq.(\ref{ccumulant}), one sees that the resultant correlation is a sum of two terms.
The first term is determined by the overall multiplicity fluctuations and the background elliptic flow.
Experimental measurements showed that the elliptic flow coefficient increases when one goes to more peripheral collisions.
It is noted that this fact plays an important role in our analysis.
The second term measures the correlations from the peripheral tube, which reflects the physics of event-by-event fluctuating IC.

We now argue that, despite its simplicity, the above analytic model captures the main characteristics of the centrality dependence of the di-hadron correlations. 
Fig.1 serves as a schematic diagram of the peripheral tube model which reproduces the main feature of the observed data. 
The parameters on the r.h.s. of Eq.(\ref{ccumulant}) are estimated as follows.
First, the multiplicity variance, $\langle N_b^2\rangle -\langle N_b\rangle ^2$, can be estimated straightforwardly through simulations of non-biased events.
For central Au+Au collisions of 0 - 10\% correspond to the impact parameter interval of 0 - 4.871 fm, and the corresponding multiplicity variance is found to be $31067(2\pi)^2$.
For peripheral collisions of 40\% - 60\% correspond to the impact parameter interval of 9.568 fm - 11.718 fm, and the multiplicity variance is found to be $7264(2\pi)^2$.
By using the above events, one also obtains the average background flow to be $v_2^{b}= 0.08$ for the central and $v_2^b=0.3$ for the peripheral window for all charged particles up to 3 GeV.
From the viewpoint of the peripheral tube model, for the most central collisions, flow harmonics such $v_2$ and $v_3$ are generated purely due to the existence of the tube, where the background is completely isotropic with $v_2^b=0$.
Therefore, one may extract information about flow harmonics $v_2^t$ and $v_3^t$ of the tube by calculations of the hydrodynamical evolutions of the events studied previously in \cite{sph-corr-4}.
By Fourier expansion, one obtains $v_2^t= 0.017$, $v_3^t= 0.015$ and $N_t=1496582(2\pi)^2$ where we assume $v_2^b=0$ and therefore $N_t=N_b$ in this case. 
The value of $N_t$ should scale proportionally to the size and number of the tubes as a function of centrality.
The latter is estimated by devising a script to calculate the volume (in terms of entropy) and the number of the tubes for different centrality windows.
Subsequently, one finds that the volume of the tube scales from $V_t\times n_t=53.7\times 3.7$ for central collisions to $V_t\times n_t=48.4\times 2.0$ for peripheral ones.
Putting all pieces together, the plots in Fig.1 are obtained by the above parameters as follows
\begin{eqnarray}
&\langle N_t^2\rangle/(2\pi)^2 =1496582\ (\hbox{central})\rightarrow 355221\ (\hbox{peripheral}),\nonumber \\
&v_2^t=0.017, v_3^t=0.015, \nonumber \\
&\langle N_b^2\rangle -\langle N_b\rangle ^2/(2\pi)^2=31067\ (\hbox{central})\rightarrow 7264\ (\hbox{peripheral}),\nonumber \\
&v_2^b=0.08\ (\hbox{central})\rightarrow 0.3\ (\hbox{peripheral}).\label{parameterset}
\end{eqnarray}
Since the first term on the r.h.s. of Eq.(\ref{ccumulant}) is mostly determined by the elliptic flow of the system, its magnitude increases from central to peripheral collisions, following that of the background elliptic flow $v_2^b$.
On the away-side where $\Delta\phi = \pi$, the contribution of the first term (shown by the red dashed lines in Fig.1b and Fig.1c) is always positive.
Consequently, for peripheral collisions, it may be just big enough to fill up the ``valley" of the second term (shown in Fig.1a), which results in a single peak on the away-side as shown by the black curve in Fig.1c.
For central collisions, on the other hand, the second term (shown in Fig.1a) dominates the overall shape of the di-hadron correlations, as one observes that the black curves in Fig.1a and Fig.1b look similar.

\begin{figure}
\begin{tabular}{ccc}
\begin{minipage}{160pt}
\centerline{\includegraphics[width=180pt]{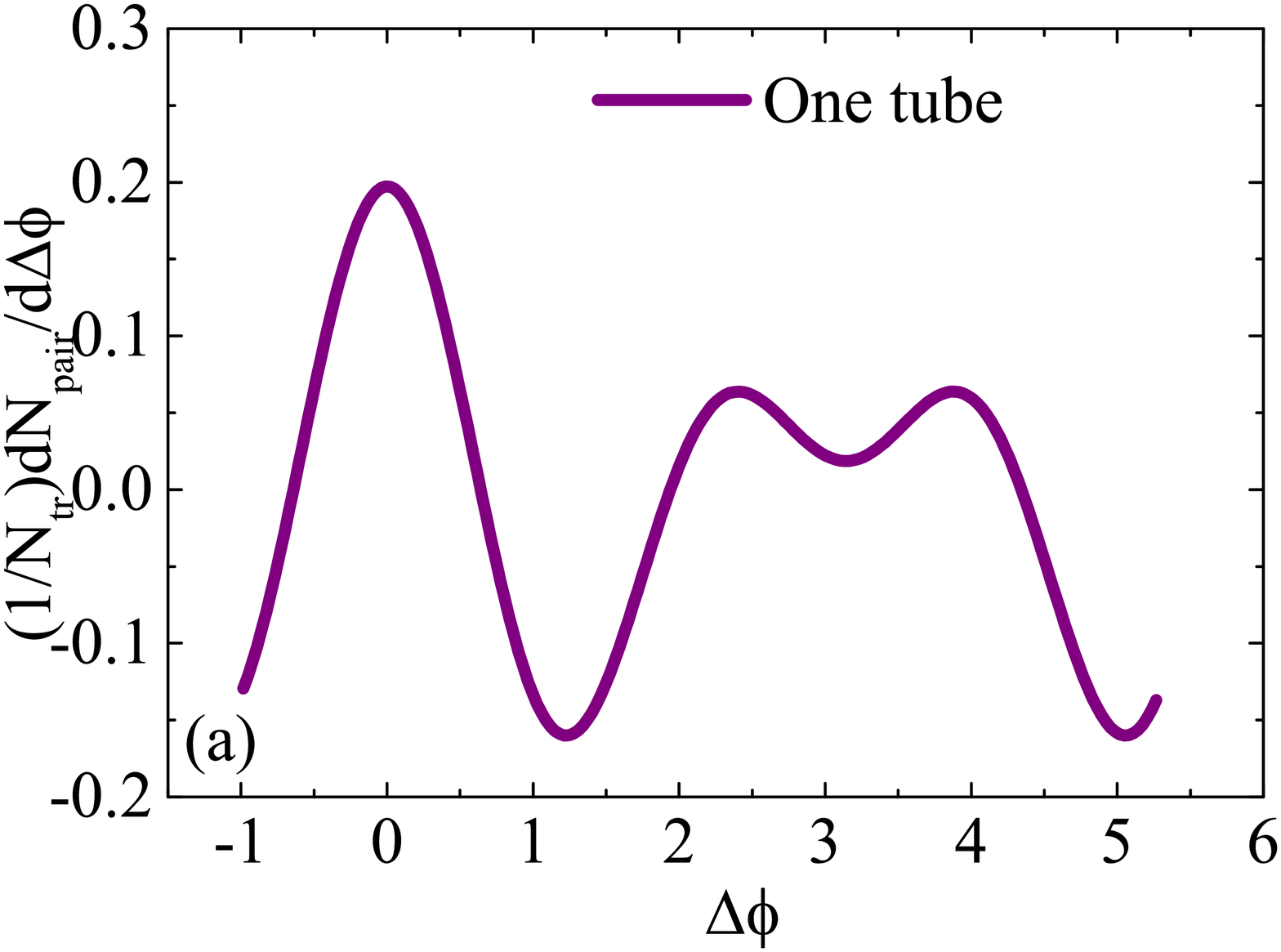}}
\end{minipage}
&
\begin{minipage}{160pt}
\centerline{\includegraphics[width=180pt]{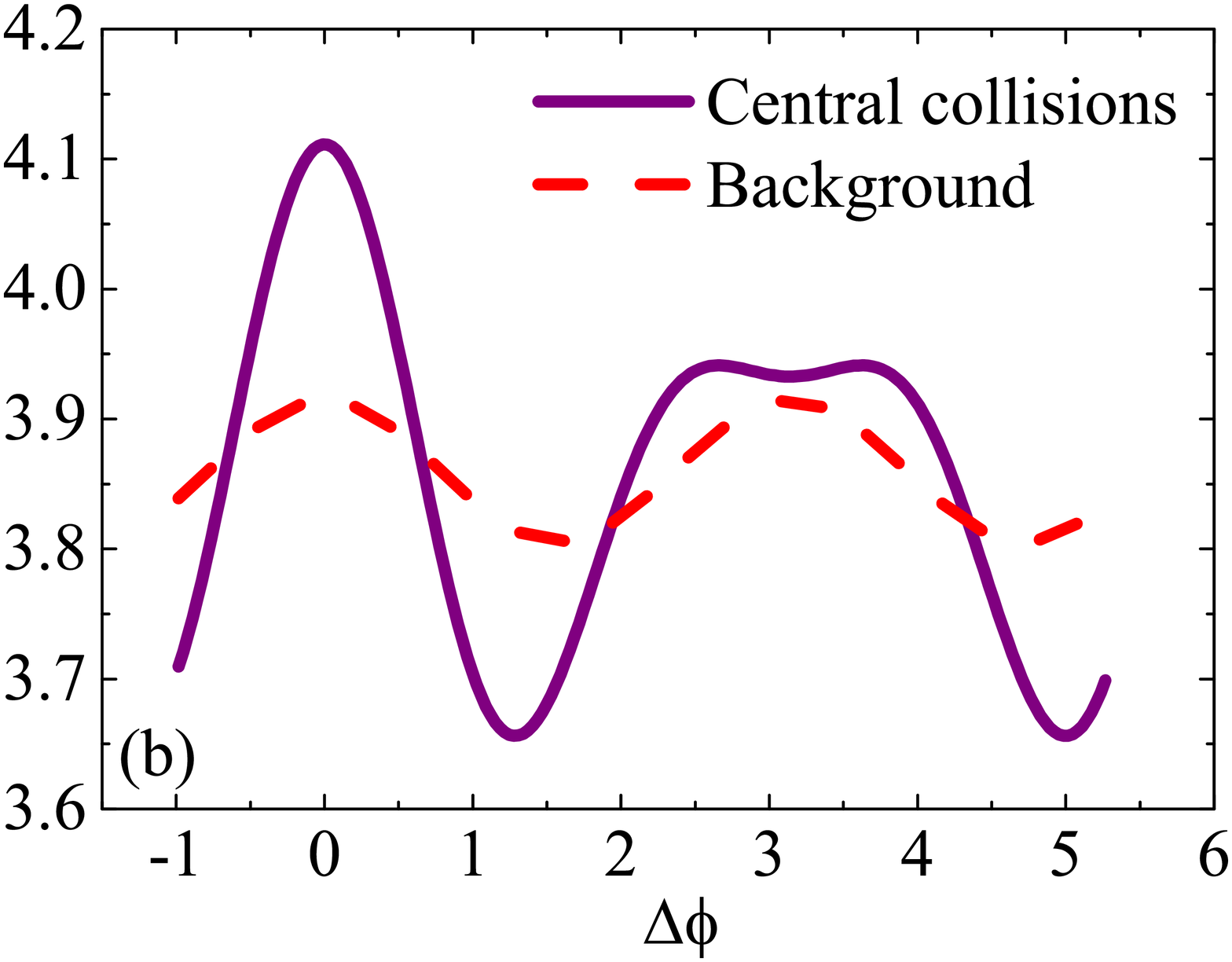}}
\end{minipage}
&
\begin{minipage}{160pt}
\centerline{\includegraphics[width=180pt]{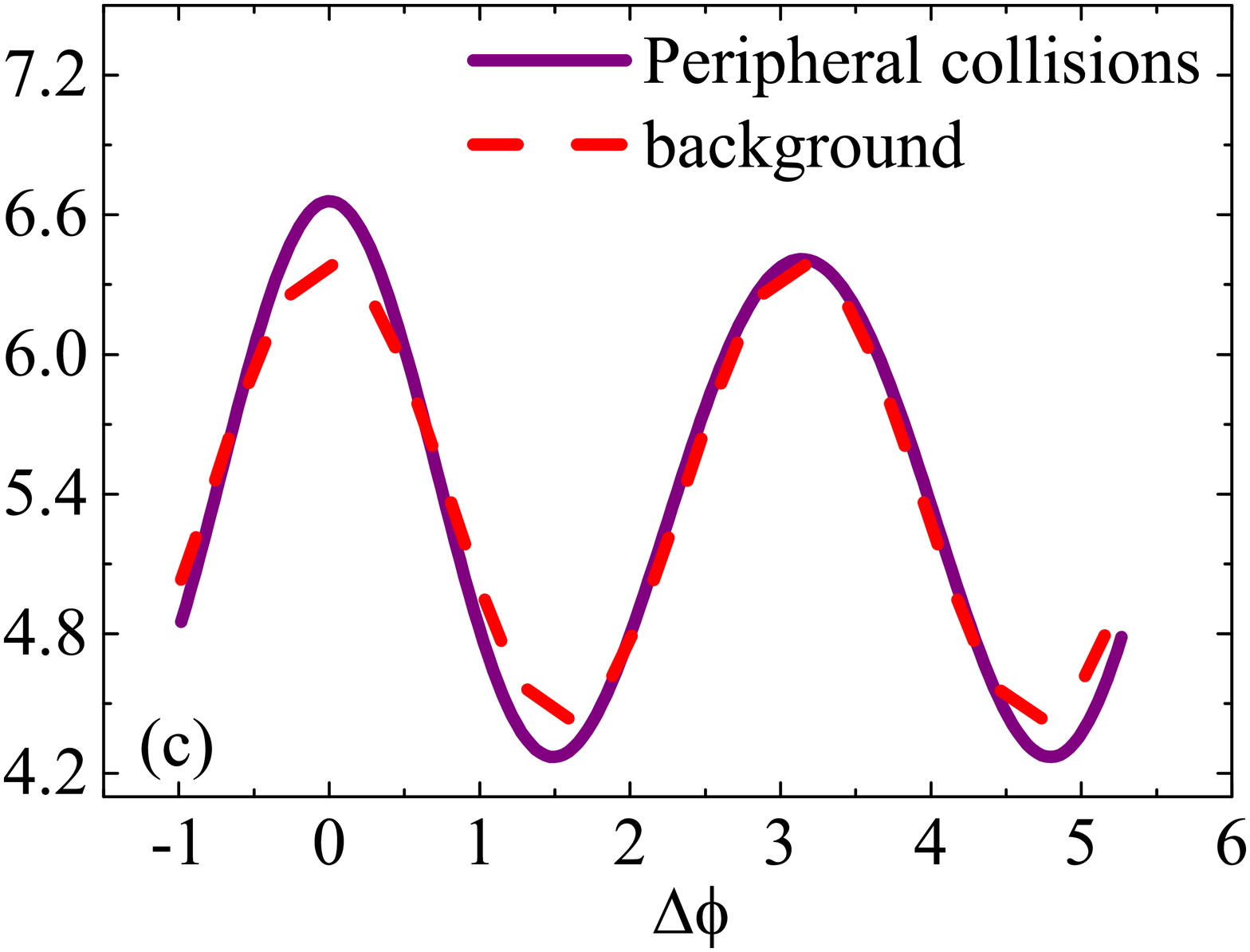}}
\end{minipage}
\end{tabular}
 \caption{(Color online) Plots of di-hadron correlations calculated by cumulant method:
 The calculations are done by using the parameters in Eq.(\ref{parameterset}), the correlation is normalized by the number of particles.
 (a) the tube contribution;
 (b) the one from the background (dashed line) and the resultant correlation (solid line) for central collisions, as
 given by Eq.(\ref{ccumulant}); and (c) the corresponding ones for the peripheral collisions.}
 \label{fig1}
\end{figure}

Now we show that very similar results will be again obtained,
if one evaluates the combinatorial mixed event contribution using ZYAM method \cite{zyam-1,zyam-2}.
The spirit of ZYAM method is to first estimate the form of
correlation solely due to the average background collective flow and,
then, to rescale the evaluated correlation by a factor $B$. The latter is determined by assuming zero signal at the minimum of the subtracted correlation.
Di-hadron correlation for the background flow is given by
\begin{eqnarray}
 \left<\frac{dN_{\text{pair}}}{d\Delta\phi}\right>^{\text{mixed}\text{(ZYAM)}}
  =B\int\frac{d\phi}{2\pi}
 \frac{dN_{bgd}}{d\phi}(\phi)\frac{dN_{bgd}}{d\phi}(\phi+\Delta\phi),
 \label{mixedz1}
\end{eqnarray}
where, according to \cite{phenix-ridge5}, the elliptic flow coefficients above are to be obtained by using event plane method.
A straightforward calculation gives
\begin{eqnarray}
 \left<\frac{dN_{\text{pair}}}{d\Delta\phi}(\phi_s)\right>^{\text{(ZYAM)}}
 &=&\frac{\langle N_b^2\rangle -B\langle N_b\rangle ^2}{(2\pi)^2}(1+2(v_2^b)^2\cos(2\Delta\phi)) \nonumber\\
 &+&(\frac{N_t}{2\pi})^2\sum_{n=2,3}2{v_n^t}^2\cos(n\Delta\phi). \label{czyam}
\end{eqnarray}
As is shown below in numerical calculations, similar results are obtained for the case of ZYAM.

\section{III. Results of numerical simulations}

In the above, our arguments are mostly based on an analytic peripheral tube model which merely considers a simplified IC.
In particular, the model only involves one peripheral tube and few most dominating flow coefficients.
The motivation of the approach is to discuss the physical insight of the problem transparently using a model with minimal parameters.
However, in realistic collisions, the IC generally contain several high energy tubes whose location, size and energy may also fluctuate.
In order to show that a hydrodynamical description indeed captures the main physical content of the observed data, we present, in the following, the results of numerical simulations by using the hydrodynamic code NeXSPheRIO.
In our calculations, the di-hadron correlations are obtained by both cumulant and the ZYAM method and compared to the data by PHENIX Collaboration.
The NeXSPheRIO code uses the IC provided by the event generator NeXuS \cite{nexus-1,nexus-rept},
and solves the relativistic ideal hydrodynamic equations with SPheRIO code \cite{sph-1st}.
By generating many NeXuS events, and solving independently the equations of hydrodynamics for each of them, one takes into account the fluctuations of IC in event-by-event basis.
At the end of the hydrodynamic evolution of each event, a Monte-Carlo generator is employed to achieve hadron emission,
in Cooper-Frye prescription, and then hadron decay is considered.
Here we note that there is no free parameter in the present simulation since the few existing ones have been fixed in earlier studies of $\eta$ and $p_T$ distributions \cite{sph-v2-3}.

Fig.2 shows the resulting di-hadron correlation by using cumulant method.
In this calculations, a total of 180 NeXuS events are generated for the centrality classes 0 - 20\% and 20\% - 40\%, and Monte-Carlo generator is invoked 1000 times for decoupling at the end of each event.
For the 60\%-92\% centrality class, 495 events are generated and each event is then followed by 800 Monte-Carlo processes.
To subtract the combinatorial background, we evaluate the two-particle cumulant.
In order to make different events similar in characters, each centrality class is further subdivided equally into smaller ones.
Then one picks a trigger particle from one event and an associated particle from a different event of the same subclass to form a hadron pair.
The azimuthal angles of the hadrons from the events of the same subclass are measured with respect to their event planes $\Psi_2$, in other words, the event planes are aligned by rotating in the transverse plane.  
Averaging over all the pairs within the same sub-centrality class, one obtains the two particle cumulant.
Background modulation is evaluated and the subtraction is done within each sub-centrality class and then they are summed up together at the end of calculation.
The numerical results are shown in Fig.2 in solid lines, where they are compared with Figs.36-38 of \cite{phenix-ridge5} in filled circles and flow systematic uncertainties in histograms.

From Fig.2, one sees that the main feature of the data is reasonably reproduced by NeXSPheRIO code.
The correlations decrease when the momentum of associated particles increase.
As one goes from the most central to peripheral collisions, the magnitude of the correlations decreases, meanwhile the away-side correlation evolves from double- to a single-peak structure. 
These features are in consistence with the data as well as with those of the peripheral tube model.
In general, it is found that the hydrodynamical simulations describe the data better for central collisions and at low and intermediate momentum range. 
For 20\% - 40\% centrality class, the results are consistent with those obtained previously \cite{sph-corr-ev-4}.
As one goes to more peripheral collisions, the simulation results underestimate the data.
Also, the discrepancy starts to increase with the transverse momentum of the associated particles.
Both of the above discrepancies are expected from a hydrodynamic simulation since in general hydrodynamic model starts to break down when dealing with either small system or large momentum. 
Another source of the discrepancy comes from the fact that the PHENIX Collaboration used the ZYAM method to evaluate the correlation instead of cumulant method adopted in Fig.1 and in \cite{sph-corr-ev-4}.
Though one may expect the main feature of the results obtained by both method to remain unchanged, it is worthwhile to carry out the calculations using exactly the same method adopted by the experimentalists. 

In the following, we calculate the correlations by using the method of PHENIX Collaboration presented in reference \cite{phenix-ridge5}.
The correlation function is then written as a sum of two terms.
\begin{eqnarray}
C(\Delta\phi) = \zeta (1+ 2\langle v_2^{trig} v_2^{asso}  \rangle \cos 2\Delta\phi) + J(\Delta\phi),
\end{eqnarray}
where $v_2^{trig}$ and $v_2^{asso}$ are the elliptic flow coefficient of trigger and associated particle respectively, determined by using event plane method \cite{phenix-v2-3}. 
The first term represents the di-hadron azimuthal correlation due to the elliptic flow up to a rescaling factor $\zeta$, which is understood to be a collective correlation presented in any event pair.
Thus the second term $J(\Delta\phi)$ measures the remaining correlations intrinsically from the proper events such as those owing to the pairs from (di)jets.
The rescaling factor $\zeta$ is fixed by the ZYAM method, which requires that the mimima of $C(\Delta\phi)$ and $\zeta (1+ 2\langle v_2^{trig} v_2^{asso}  \rangle \cos 2\Delta\phi)$ attain the same value, or in other words, the subtracted correlation $J(\Delta\phi)$ assumes zero at its minimum.
\begin{eqnarray}
C(\Delta\phi_{\text{min}}) = \zeta (1+ 2\langle v_2^{trig} v_2^{asso}  \rangle \cos 2\Delta\phi_{\text{min}}),
\end{eqnarray}
or
\begin{eqnarray}
J(\Delta\phi_{\text{min}}) = 0 .
\end{eqnarray}
Here $\zeta$ carries the same physica content as $B$ in Eq.(\ref{mixedz1}) up to a normalization factor.

To faithfully reproduce the experimental procedure, following the hydrodynamic evolution of each random IC, the Monte-Carlo hadron generator is invoked 200 times for each event.
In order to obtain better statistics, a total of 3,500 events are generated for the 0 - 20\% centrality class, 2,000 events for the 20\%-40\% centrality class and 5,000 events for the 60\%-92\% centrality class.
The elliptic flow coefficients for the trigger and associated particles are obtained using the event plane method \cite{phenix-v2-3} together with corresponding acceptance cuts adopted by PHENIX Collaboration.
To evaluate the event plane, one considers hadrons within the pseudo rapidity window $|\eta| < 1$ and with transverse momentum $p_T > 0.1$ GeV.
The elliptic flow is evaluated by taking into account hadrons within the pseudo rapidity range $|\eta| < 1 $.
We further approximate $\langle v_2^{trig} v_2^{asso}  \rangle = \langle v_2^{trig} \rangle \langle v_2^{asso}  \rangle$. 
The resulting $v_2\{\text{EP}\}$ are shown in the Table I.
\begin{table}[!hbt] 
\caption{$v_2\{\text{EP}\}\pm \Delta v_2$ for transverse momentum ranges of interest} 
\medskip
\begin{tabular}{c|c|c|c}
\hline\hline
$p_T$ range (GeV) & $ 0 - 20\% $ & $ 20\% - 40\% $ & $ 60\% - 92\% $\\ 
\hline
$0.4 - 1.0 $ & $0.0399 $ & $0.0729 $ & $0.0739 $ \\
$1.0 - 2.0 $ & $0.0858 $ & $0.1543 $ & $0.1460 $ \\
$2.0 - 3.0 $ & $0.1387 $ & $0.2466 $ & $0.2369 $ \\
\hline
\end{tabular}
\label{tb1}
\end{table}

The resulting di-hadron correlations are shown in Fig.3.
Again, one finds that the hydrodynamic calculations describe the data better for central collision and at small transverse momentum.
Although smaller than in the cumulant method, as one goes to more peripheral windows as well as higher momentum range, deviations appear.
However, the main features of the data are reasonably reproduced by the calculations which imply that hydrodynamic model captures the main physics in the observed centrality dependence of di-hadron correlations.

\section{IV. Concluding remarks}

The harmonic coefficients quantify the inhomogeneity of the energy density in fluctuating IC, but they are not the only way to achieve such a measurement.
In fact, since the flow harmonics are obtained by the event average of particle correlation functions, some particular information on the characteristic of each individual event might be averaged out during the process.

In our picture, the focus is given to the individual events.
Unlike flow harmonics, which can be obtained by energy density distribution of IC without any protruding peak, the physical content of the present model is not associated with those fluctuations whose wave-length is comparable to the system size.  
Instead, the inhomogeneity is expressed in terms of localized peripheral hot tubes, which belong to each individual event. 
Therefore, the evolution of the shape of the away-side structure in two-particle correlation from central to peripheral collisions is not attributed to the harmonic coefficients, but to the contributions produced by the peripheral tube (hotspot) and those by background flow and its fluctuations.
Although one can always decompose particle distribution and correlation in terms of harmonic coefficients by using Fourier expansion, we show that the peripheral tube model offers an alternative viewpoint where the emergence of observed di-hadron correlation can be understood in terms of a simple physical mechanism.  
The present approach provides an intuitive explanation of the centrality as well as event plane dependence of the observed evolution of the away-side correlations.

To summarize, we argued that the observed centrality dependence of di-hadron correlations can be understood in terms of the peripheral tube model, where one assumes the superposition of the centrality-dependent background flow and a small portion of deflected flow due to the presence of a peripheral tube.
In our simple analytic model, the observed features in PHENIX data can be reproduced by a proper choice of parameters.
As was discussed previously \cite{sph-corr-ev-4}, the peripheral tube model gives a unified description of the ``ridge'' structure, both for the near-side and the away-side ones.
In this interpretation, these structures in the correlation are causally connected, their appearance does not dependent on any global structure of the IC.
In other words, the anisotropic parameters are rather related to the existence of spiky localized hot spots in the IC than to the geometrical form of a smooth energy distribution as commonly imagined.
The importance of such granularities in IC for the anisotropic flow $v_2$ has been proposed in \cite{sph-v2-2}, but here we pointed out that for the triangle flow, their roles become more explicit.   
Ongoing studies on the event planes correlation as well as symmetric cumulant \cite{atlas-vn3,atlas-vn5,alice-vn5} possibly provide a way to distinguish between different models.
Therefore, the generalization of the present approach to understand the above measurements with respect to the collision geometry is an interesting topic, and we plan to carry out such analysis in the near future.

\section*{Acknowledgments}
We are thankful for valuable discussions with Rafael Derradi, Jun Takahashi, Fuqiang Wang, Che Ming Ko, Paul Sorensen and Jiangyong Jia.
We gratefully acknowledge the financial support from 
Funda\c{c}\~ao de Amparo \`a Pesquisa do Estado de S\~ao Paulo (FAPESP), 
Funda\c{c}\~ao de Amparo \`a Pesquisa do Estado de Minas Gerais (FAPEMIG), 
Funda\c{c}\~ao de Amparo \`a Pesquisa do Estado do Rio de Janeiro (FAPERJ),
Conselho Nacional de Desenvolvimento Cient\'{\i}fico e Tecnol\'ogico (CNPq),
and Coordena\c{c}\~ao de Aperfei\c{c}oamento de Pessoal de N\'ivel Superior (CAPES).
F. G. Gardim would like to acknowledge CNPq project under Contract No. 449694/2014-3.

\begin{figure}
  \centerline{\includegraphics[width=16.cm]{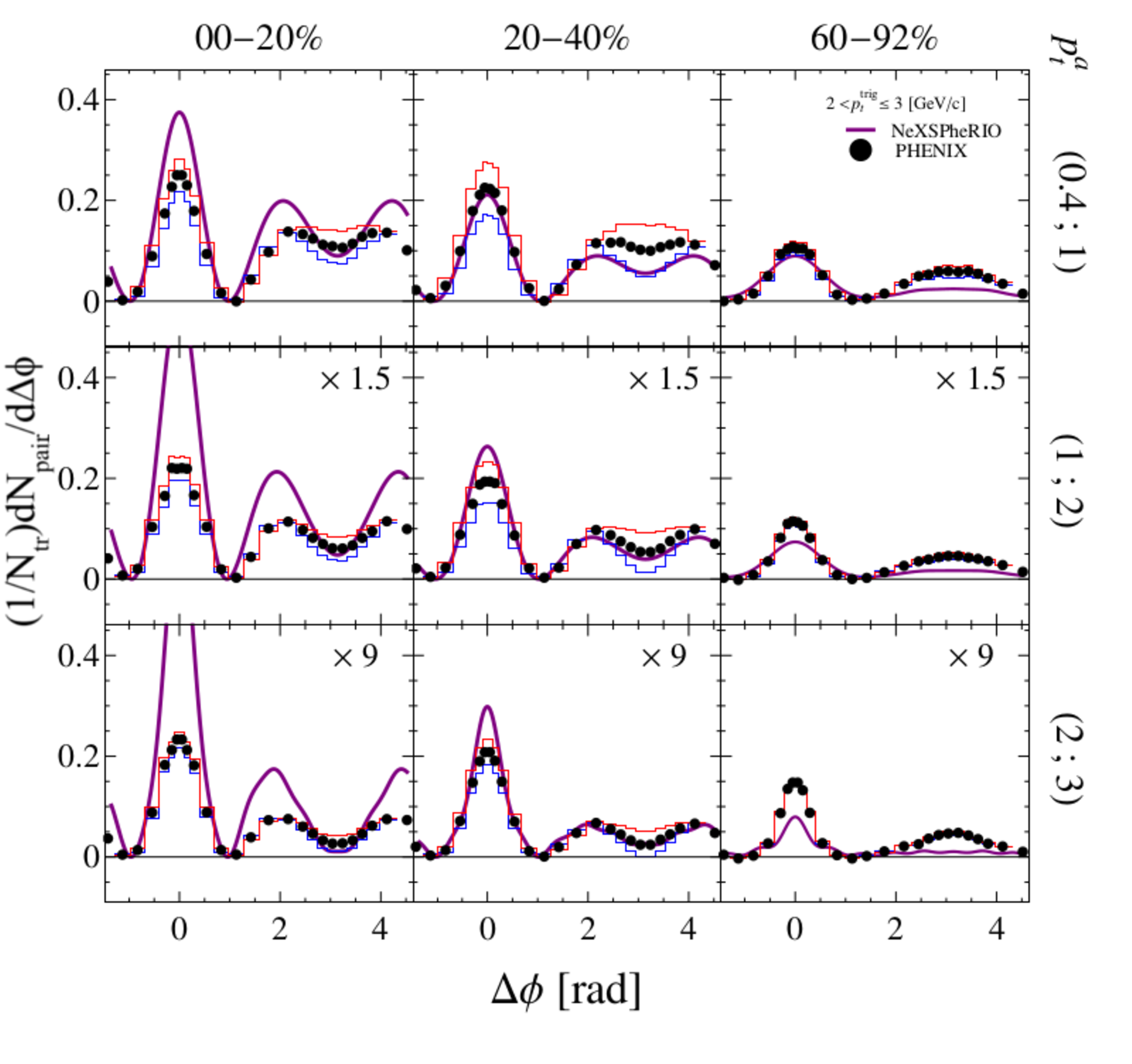}}
  \caption{(Color online) The subtracted di-hadron correlations as a function of $\Delta\phi$ for different centrality windows and $p_T^{a}$ range for 200A GeV Au+Au collisions.
   NeXSPheRIO results in solid curves, are compared with PHENIX data in filled circles. }
  \label{fig2}
\end{figure}

\begin{figure}
 \centerline{\includegraphics[width=16.cm]{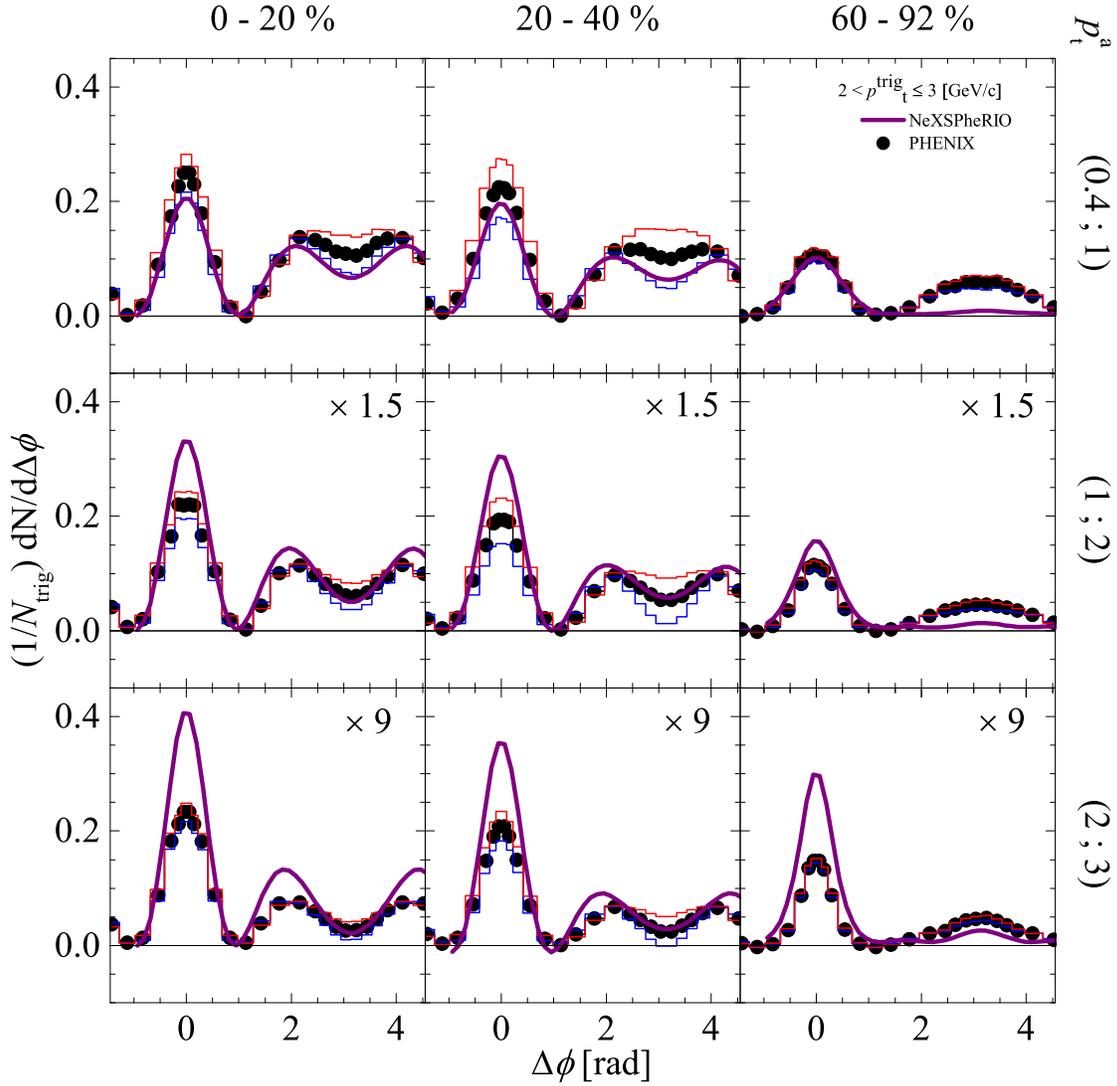}}
 \caption{(Color online) The same as Fig.2 but obtained by using ZYAM method described in the text.}
 \label{fig3}
\end{figure}

\bibliographystyle{h-physrev}
\bibliography{references_qian}{}

\end{document}